\documentclass[osajnl,preprint,twocolumn,superscriptaddress,10pt]{revtex4-1}
\usepackage{amsmath}
\usepackage{bigints}
\usepackage{graphicx}

\begin{document}

\title{No free lunch: the tradeoff between heralding rate and efficiency in microresonator-based heralded single photon sources}

\author{Z. Vernon}
\email{zachary.vernon@utoronto.ca}
\affiliation{Department of Physics, University of Toronto, 60 St. George Street, Toronto, Ontario, Canada, M5S 1A7}
\author{M. Liscidini}
\affiliation{Department of Physics, University of Pavia, Via Bassi 6, Pavia, Italy}
\author{J.E. Sipe}
\affiliation{Department of Physics, University of Toronto, 60 St. George Street, Toronto, Ontario, Canada, M5S 1A7}

\date{\today}

\begin{abstract}
Generation of heralded single photons has recently been demonstrated using spontaneous four-wave mixing in integrated microresonators. While the results of coincidence measurements on the generated photon pairs from these systems show promise for their utility in heralding applications, such measurements do not reveal all of the effects of photon losses within the resonator. These effects, which include a significant degradation of the heralding efficiency, depend strongly on the relative strengths of the coupling of the ring modes to loss modes and channel modes. We show that the common choice of critical coupling does not optimize the rate of successfully heralded photons, and derive the coupling condition needed to do so, as well as the condition needed to maximize the rate of coincidence counts. Optimizing these rates has a considerable negative effect on the heralding efficiency.
\end{abstract}

\maketitle

Heralded single photons are an important resource both for optical quantum information processing, and for fundamental investigations of nonlinear quantum optics at the single photon level. There is particular interest in developing monolithically integrated sources of heralded single photons, which would enable their use in an on-chip optical setting. Several such implementations have recently been demonstrated using spontaneous four-wave mixing (SFWM) in integrated microresonators \cite{Davanco2012,He2015,Reimer2014,Savanier2015}. 

In typical implementations each photon of a generated pair is emitted into one of two distinct optical modes: a heralding mode (HM) carries a herald photon, the presence of which then indicates the existence of a single photon in an output mode (OM). These modes might correspond to the signal and idler fields generated by SFWM in a microresonator \cite{Davanco2012,Reimer2014,Helt2010}, or to the two outputs of a degenerate photon pair splitter \cite{He2015}. By discarding experimental runs in which no heralding photon is detected in the HM, the experimenter effectively post-selects on those runs in which a single photon is present in the OM. In an ideal device, detection of an HM photon would \emph{guarantee} the existence of an OM photon. In practice, even assuming perfect detection efficiency, photon losses within the photon pair source lead to events in which the herald photon is detected but the OM photon is lost. After detecting the herald, the OM mode cannot therefore be represented as the desired pure state $\rho_\mathrm{OM}=\vert 1\rangle \langle 1 \vert$, but rather $\rho_\mathrm{OM}=P_\mathrm{vac}\vert 0 \rangle \langle 0 \vert + P_\mathrm{1}\vert 1 \rangle \langle 1 \vert$, where $P_\mathrm{vac}$ corresponds to the probability of losing the OM photon given the successful generation of a photon pair and subsequent detection of the HM photon, and $P_1=1-P_\mathrm{vac}$. The existence of this vacuum probability degrades the utility of the heralding device: if $P_\mathrm{vac}$ is significant the very purpose of the herald is compromised. Note that this narrative neglects the effects of spectral correlation between the signal and idler photons, which would lead to the one-photon state itself being expressed not as $\vert 1\rangle\langle 1\vert$, but as a mixture of states involving a photon at different frequencies. Discussions of heralded state purity in photon pair sources usually focus on such spectral effects, often neglecting the additional mixedness arising from photon losses. For the purposes of our discussion the density operator $\rho_\mathrm{OM}$ is understood to be decomposed only in terms of the vacuum and \emph{total} one-photon probabilities. We assume the system is pumped by a pulse with bandwidth comparable to that of the resonator modes, almost eliminating the impurity in the heralded OM state that arises from spectral correlations between the OM and HM photons \cite{Helt2010,Vernon2015}.
\begin{figure}
\includegraphics{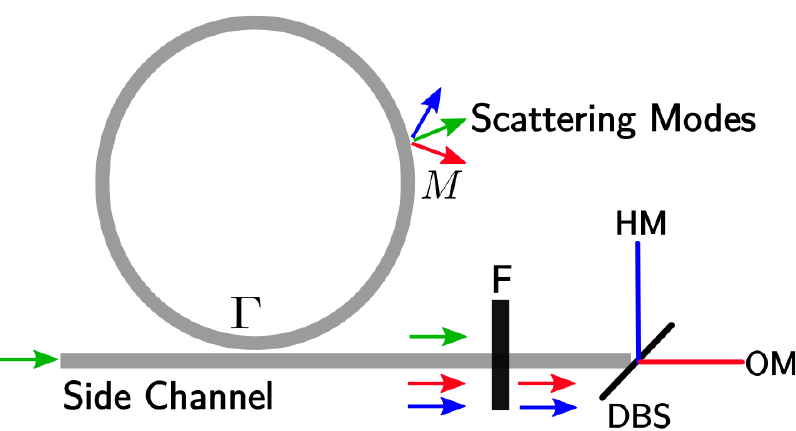}\caption{\textbf{Schematic of ring-channel system for generating heralded single photons.} Input pump (green) generates signal and idler photons (blue and red, respectively) within the ring, which are then either lost or exit to the side channel. A filter (F) removes the pump component, and a dichroic beamsplitter (DBS) separates the signal and idler for individual detection. These elements are included for illustrative purposes only; losses due to them, which would affect any source, are neglected. The signal serves as the heralding mode (HM), while the idler is the output mode (OM). The coupling rate to the side channel is given by $\Gamma$; the effective loss rate from the ring modes into scattering modes is given by $M$. As described in the text, these rates can be related in a simple way to the extrinsic and intrinsic quality factors of the ring.} \label{ringchannel}
\end{figure}

The effects of losses cannot safely be ignored. In this Letter, we show that microresonator-based heralded single photon sources suffer from a tradeoff between the heralding efficiency and the heralding rate. We focus on a specific popular microresonator structure, that of a ring resonator fabricated from a material possessing a nonlinear optical response and coupled to a channel waveguide. But we emphasize that our conclusions do not critically depend on which specific resonator structure is assumed.  We consider two heralding strategies: one in which two pump photons of identical frequency generate a pair of photons with well separated frequencies \cite{Clemmen2009,Azzini2012,Davanco2012,Reimer2014} as shown in Fig. \ref{ringchannel}, as well the recently demonstrated strategy \cite{He2015} where two pump photons of differing frequencies produce a degenerate photon pair with identical frequencies, which is then split using the time-reversed Hong-Ou-Mandel effect resulting in the output of two single photons of identical frequency into physically distinct channels. While for clarity we develop our discussion in reference to the former strategy, our results are equally applicable to the degenerate pair splitter implementation.

We recently studied \cite{Vernon2015} the effects of photon losses in the microring structure shown in Fig. \ref{ringchannel}, modeling the effects of scattering loss with the inclusion of a second ``phantom channel" into which ring photons can be lost. In particular, at sufficiently low pump powers such that generation of multiple photon pairs is suppressed, the ratio $r$ between the probabilities of detecting a signal (in this context taken to be the HM) photon whose idler (taken to be the OM) pair partner was lost, and detecting a signal photon whose pair partner is present, was found to be
\begin{eqnarray}\label{singles-coincidences-ratio}
r=\frac{M}{\Gamma},
\end{eqnarray}
where $M$ is the effective coupling rate of the ring modes to scattering modes, and $\Gamma$ the coupling rate between the ring modes and the side channel. These coupling rates are related in a simple way to the more experimentally accessible quality factor $Q$ of the resonator. This quality factor can be decomposed into intrinsic and extrinsic contributions via
\begin{eqnarray}
\frac{1}{Q}=\frac{1}{Q_\mathrm{int}} + \frac{1}{Q_\mathrm{ext}},
\end{eqnarray}
where the intrinsic contribution $Q_\mathrm{int}$ is defined to be the quality factor of the resonator were it completely isolated and uncoupled from the side channel, so that the extrinsic contribution $Q_\mathrm{ext}$ arises only from the coupling to the the side channel. The ring-channel coupling rate $\Gamma$ is then given by
\begin{eqnarray}
\Gamma = \frac{\omega}{Q_\mathrm{ext}},
\end{eqnarray}
where $\omega$ is the angular frequency of the relevant ring mode, while the loss coupling rate $M$ is given by
\begin{eqnarray}
M = \frac{\omega}{Q_\mathrm{int}}.
\end{eqnarray}

The vacuum and one-photon probabilities $P_\mathrm{vac}$ and $P_1$ of the heralded OM state $\rho_\mathrm{OM}=P_\mathrm{vac}\vert 0\rangle\langle 0\vert + P_1\vert 1 \rangle\langle 1 \vert$ satisfy $P_\mathrm{vac}/P_1=r$; normalization of $\rho_\mathrm{OM}$ then gives
\begin{eqnarray}\label{rho_OM}
\rho_\mathrm{OM}=\frac{M}{\Gamma+M}\vert 0 \rangle\langle 0\vert + \frac{\Gamma}{\Gamma+M}\vert 1\rangle\langle 1\vert.
\end{eqnarray}
In terms of the quality factors, $\rho_\mathrm{OM}$ can also be written as
\begin{eqnarray}
\rho_\mathrm{OM} = \frac{Q}{Q_\mathrm{int}}\vert 0\rangle\langle 0\vert + \frac{Q}{Q_\mathrm{ext}}\vert 1\rangle\langle 1 \vert.
\end{eqnarray}
While discussions of heralded state purity for SFWM-based photon pair sources often focus on spectral effects, for our purposes we assume the system has been pumped such that the one-photon probability in $\rho_\mathrm{OM}$ is represented by a pure state; any mixedness in $\rho_\mathrm{OM}$ then arises solely from photon losses in the resonator. The purity of the heralded OM state is thus a function of the coupling rates, and is given by 
\begin{eqnarray}\label{purity}
\mathrm{Tr}(\rho_\mathrm{OM}^2)=\frac{\Gamma^2 + M^2}{(\Gamma+M)^2}.
\end{eqnarray}
At low pump powers, and for a pump pulse with bandwidth on the order of $(\Gamma + M)$ so that spectral purity is indeed not an issue, the overall rate $J_\mathrm{HM}$ of outgoing available photons in the HM was found to scale with $\Gamma$ and $M$ as \cite{Vernon2015,Vernon2015b}
\begin{eqnarray}
J_\mathrm{HM}=\beta\frac{\Gamma^3}{(\Gamma+M)^5},
\end{eqnarray}
where $\beta=4\alpha\Lambda^2\mathcal{E}_\mathrm{pulse}^2f_P/(\hbar\omega_P\pi)^2$, with $\Lambda$ a constant related to the third-order nonlinear optical response in the ring, $\omega_P$ the frequency of the pump mode, $\mathcal{E}_\mathrm{pulse}$ the total energy of each pump pulse in the channel and $f_P$ the repetition rate of the pump laser; $\beta$ is independent of any coupling rates. The dimensionless factor $\alpha$ depends on the input pulse profile, and is given by
\begin{eqnarray}\label{alpha_defn}
\alpha=\int dq\int dq' \frac{\bigg\vert \int dp \frac{D(p)D(q+q'-p)}{(-ip+1)(-i(q+q'-p)+1)}\bigg\vert^2}{(q^2+1)(q'^2+1)},
\end{eqnarray}
where $D(\omega/(\Gamma+M))$ is the normalized input pulse amplitude profile in the frequency domain, which we take to have a characteristic width $\Delta \omega$ on the order of $\Gamma + M$; for a Gaussian profile $D(p)=(2/\pi)^{(1/4)}e^{-p^2}$ a numerical estimate yields $\alpha\approx 1.6$.
The scaling of $J_\mathrm{HM}$ for a fixed $\mathcal{E}_\mathrm{pulse}$ and $f_P$ with these rates arises from several underlying physical processes: 
\iftrue (i) for a pair to be produced and a signal photon emitted into the channel, two pump photons must enter the ring from the channel, and one photon must enter the channel from the ring, giving rise to three factors of $\Gamma/(\Gamma+M)$, (ii) the field enhancement of each mode in the ring is proportional to $(\Gamma+M)^{-1}$ \cite{Helt2012b}; since four modes participate in SFWM (pump, pump, signal, idler) this yields $(\Gamma+M)^{-4}$, and (iii) the input pump power is proportional to $(\Gamma+M)$, which we take as the pump bandwidth to suppress spectral entanglement; the quadratic dependence of the pair generation probability on the input power thus gives rise to an additional factor of $(\Gamma+M)^2$

The loss rate $M$ is a function of the quality of the fabrication process and is not easily controlled. On the other hand, the coupling rate $\Gamma$ can be controlled by constructing systems with varying distances between the ring and the channel at the coupling point \cite{Heebner2008}. Considering the variation of $J_\mathrm{HM}$ with respect to $\Gamma$ for fixed $M$ and $\beta$, we find that the HM photon flux is maximized when $\Gamma=M$. This regime in which the ring-channel coupling rate equals the loss rate is referred to as \emph{critical coupling}, and represents the best choice of $\Gamma$ if the only goal is to maximize the outgoing available photon flux. Operating at critical coupling also maximizes the intraring pump photon number $N_P$ for a given input power, and is therefore a typical choice in experiments. As is clear from Eq. (\ref{purity}), however, this choice has serious consequences for the purity of the state of the OM; at critical coupling we find $\mathrm{Tr}(\rho_\mathrm{OM}^2)=0.5$, which is precisely the global \emph{minimum} of the heralded state purity, since when $\Gamma=M$ the OM photon is lost exactly as often as it is present. Maximizing the purity of $\rho_\mathrm{OM}$ would suggest taking $\Gamma \gg M$, since $\mathrm{Tr}(\rho_\mathrm{OM}^2)\to 1$ in that limit. Yet operating in this over-coupled regime leads to suppression of the flux of available HM photons $J_\mathrm{HM}$, as $J_\mathrm{HM}\sim \Gamma^{-2}$  for $\Gamma \gg M$.

These results suggest that, depending on the specific application, microresonator-based heralded single photon generators should not necessarily be designed to operate at critical coupling. Indeed, while critical coupling does optimize the flux of HM photons for a fixed input power, it does not maximize the available rate of ``successful heralds,'' those being HM photon detection events wherein the OM photon is present. This successful heralding rate $J_\mathrm{heralds}$ is given by the product of the intraring pair generation rate $J_\mathrm{ring}=\beta \Gamma^2/(\Gamma+M)^4$ with the fraction of pairs which exit to the side channel,
\begin{eqnarray}\label{heralding_rate}
J_\mathrm{heralds}=J_\mathrm{ring}\times \frac{\Gamma^2}{(\Gamma+M)^2}=\beta\frac{\Gamma^4}{(\Gamma+M)^6},
\end{eqnarray}
and is maximized at any given input power by choosing $\Gamma=2M$, yielding a heralded OM state $\rho_\mathrm{OM}=(1/3)\vert 0\rangle\langle 0\vert + (2/3)\vert 1\rangle\langle 1\vert$ which has corresponding purity $\mathrm{Tr}(\rho_\mathrm{OM}^2)\approx 0.56$. Thus this choice of $\Gamma$ above critical coupling to maximize the heralding rate also improves the purity, albeit very slightly. For applications wherein the successful heralding rate is of principal importance, the ring-channel should therefore be operated in this moderately over-coupled regime. This would be the best choice for experiments which post-select on coincidences in the HM and OM modes, rendering unsuccessful heralds unimportant \cite{Silverstone2014}.

Having calculated the heralding rate $J_\mathrm{heralds}$, as well as the total rate $J_\mathrm{HM}$ of available photons in the HM, we can calculate the heralding efficiency $\eta$, which is defined as the ratio between these two rates:
\begin{eqnarray}\label{efficiency}
\eta \equiv \frac{J_\mathrm{heralds}}{J_\mathrm{HM}} = \frac{\Gamma}{\Gamma + M}.
\end{eqnarray}
This is precisely the one-photon probability present in the heralded output state $\rho_\mathrm{OM}$; comparison with Eq. (\ref{rho_OM}) allows us to write $\rho_\mathrm{OM}$ as
\begin{eqnarray}
\rho_\mathrm{OM}=(1-\eta)\vert 0\rangle \langle 0 \vert + \eta \vert 1\rangle\langle 1\vert.
\end{eqnarray}
The purity is thus directly related to the efficiency $\eta$ as $\mathrm{Tr}(\rho_\mathrm{OM}^2)=2\eta^2-2\eta+1$. While a heralding efficiency as close as possible to unity is desirable, the choice of critical coupling to maximize $J_\mathrm{HM}$ gives $\eta=0.5$, whereas choosing $\Gamma=2M$ to maximize $J_\mathrm{heralds}$ yields $\eta\approx 0.67$, only a modest improvement to the critically coupled regime. As with the purity, maximizing the heralding efficiency requires a strongly over-coupled system with $\Gamma \gg M$. 

By inverting Eq. (\ref{efficiency}) to obtain $\Gamma$ as a function of $\eta$ and $M$, we can express the heralding rate as a function of heralding efficiency:
\begin{eqnarray}\label{defn_J_vs_eta}
J_\mathrm{heralds}(\eta) = \beta\frac{\eta^4(1-\eta)^2}{M^2}.
\end{eqnarray}
The prefactor $\beta/M^2$ depends on the scattering rate $M$ and, through its dependence on $\beta$, on the nonlinear coefficient $\Lambda$ as well as the pump pulse energy and pump pulse profile. As an example we compare the performance of silicon and silicon nitride rings, based on current fabrication capabilities: Silicon nitride microrings can be constructed with intrinsic quality factors on the order of $10^6$, about 10 to 100 times greater than the best silicon microrings; however, the latter have a nonlinear coefficient $\Lambda$ approximately 1000 times larger than that of silicon nitride \cite{Vernon2015b,He2015,Dutt2015}. Using appropriate parameters in the expression (\ref{defn_J_vs_eta}), we plot that expression for silicon and silicon nitride rings in Fig. \ref{JvsEta}. There is clearly a tradeoff between heralding rate and heralding efficiency, with the heralding rate for each ring dropping off sharply as the desired heralding efficiency exceeds 0.67. Because $\Lambda$ for silicon is larger than that of silicon nitride, the heralding rate for a silicon-based system would be hundreds of times greater than that of a silicon nitride-based system at the same efficiency $\eta$. Note that in the plot we have taken $\alpha=1.6$ for both systems, at all $\eta$, corresponding to a pump pulse bandwidth on the order of $\Gamma+M$; otherwise spectral correlations would arise. But it is crucial to recall that increasing $\eta$ (\ref{efficiency}) for fixed $M$ requires an increase in the coupling rate $\Gamma$ of the channel to the ring, and thus requires a corresponding increase in the pump pulse bandwidth. Pump dispersion in the driving channel then can limit device performance. Somewhat arbitrarily, we set 10 ps as the minimum allowable pulse duration. This leads to a limit of a heralding efficiency of approximately 0.5 for typical silicon rings and 0.98 for those with the highest reported intrinsic quality factors; for the best silicon nitride rings the superior quality factor permits $\eta$ to be as high as 0.998. These values are indicated by the red diamonds on each curve in Fig. (\ref{JvsEta}); points to the right of these markers require a pump pulse duration below 10 ps.
\begin{figure}[h]
\includegraphics[width=1\columnwidth]{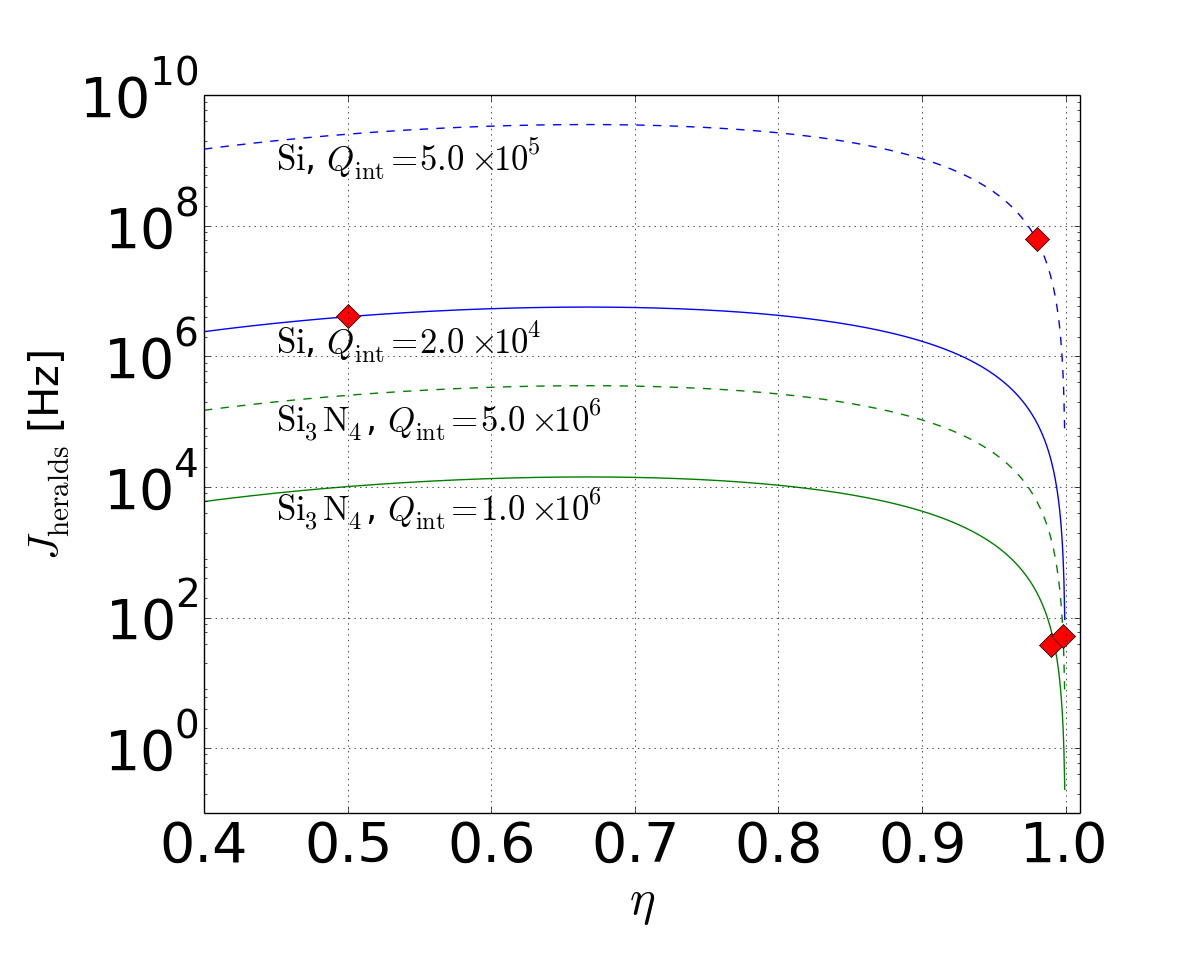}
\caption{\textbf{Heralding rate vs. heralding efficiency.} Heralding rate $J_\mathrm{heralds}$ as a function of heralding efficiency $\eta$ for typical silicon (upper blue curves) and silicon nitride (lower green curves) microring resonator systems. The red diamonds on each curve indicate the point at which the required pump pulse duration is reduced to 10 ps, representing a practical approximate upper limit for $\eta$ in each system; points to the right of these markers will suffer significantly from dispersion of the pump pulse as it propagates toward the ring. The nonlinear parameter for silicon (silicon nitride) was taken to be $\Lambda=10$ kHz (10 Hz). Solid curves represent rings with typical intrinsic quality factors, while the dashed curves correspond to estimates of the highest intrinsic quality factors reported in the literature \cite{Savanier2015,Okawachi2011}. For all systems the pump was taken to have a Gaussian frequency profile with central wavelength 1550 nm in vacuum and bandwidth on the order of $\Gamma+M$ for each $\eta$. The pump energy per pulse was taken as $\mathcal{E}_\mathrm{pulse}=10$ pJ with repetition rate $f_P=1$ MHz, giving a total average pump power of 10 $\mu$W.}\label{JvsEta}
\end{figure}

Aside from heralding rate and efficiency concerns, for certain applications one might prioritize the \emph{coincidence rate} of simultaneously detected HM-OM photon pairs. For the microresonator devices under consideration, this coincidence rate $J_\mathrm{coincidences}$ can be calculated from the appropriate Glauber formula $\mathrm{Avg}\left[\langle \psi_\mathrm{HM}^\dagger(t) \psi_\mathrm{OM}^\dagger(t') \psi_\mathrm{HM}(t) \psi_\mathrm{OM}(t')\rangle\right]$, where $\psi_{J}(t)$ are the Heisenberg operators for the output fields $J=\{\mathrm{HM},\mathrm{OM}\}$ at time $t$, and $\mathrm{Avg}[\cdot]$ denotes a time average over $t'$ within a small time window $\delta t$ about $t$. The coincidence rate is found to scale with the coupling rates as\cite{Vernon2015,Vernon2015b} 
\begin{eqnarray}\label{coinc_rate}
J_\mathrm{coincidences}\propto \frac{\Gamma^4}{(\Gamma+M)^5}.
\end{eqnarray}
The coincidence rate differs by a factor of $\Gamma + M$ from the successful heralding rate $J_\mathrm{heralds}$. This is to be expected, since a coincidence detection is sensitive to the average lifetime $(\Gamma+M)^{-1}$ of the photons in the ring; the probability of simultaneous detection of a pair of simultaneously generated photons is inversely proportional to this lifetime, as the OM photon may exit the resonator, and therefore arrive at the detector, some time earlier or later than the corresponding HM photon. \emph{Simultaneous} in this context means joint detection within a small time window $\delta t$ such that $\delta t \ll (\Gamma+M)^{-1}$, which we assumed in deriving (\ref{coinc_rate}); if $\delta t$ is increased to significantly exceed the photon lifetime, the coincidence rate will scale with the coupling rates in exactly the same way as $J_\mathrm{heralds}$. For small $\delta t$ this coincidence rate (\ref{coinc_rate}) is maximized at fixed input power when  $\Gamma$ is chosen to satisfy $\Gamma=4M$. The corresponding heralded output state for a system engineered to optimize the coincidence rate is then given by $\rho_\mathrm{OM} = (1/5)\vert 0\rangle\langle 0 \vert + (4/5)\vert 1\rangle\langle 1\vert$, with a purity $\mathrm{Tr}(\rho_\mathrm{OM}^2)\approx 0.68$; the heralding efficiency in this case is $\eta = 0.80$.

Our discussion so far has addressed the scaling relationship of various generation rates with the coupling rates $\Gamma$ and $M$ for \emph{fixed} input pump power. In each case increasing $\Gamma$ above $M$, which is necessary in order to increase the heralding efficiency and obtain heralded OM states with higher purity, degrades the corresponding generation rate. It is natural to ask whether the degradation in the heralding rate $J_\mathrm{heralds}$ can be compensated for by simply increasing the pump power. In principle this is possible: the factor $\beta$ in Eq. (\ref{heralding_rate}) is quadratic in the energy of the input pulse $\mathcal{E}_\mathrm{pulse}$, and can be scaled appropriately to maintain the desired heralding rate as $\Gamma$ is increased. However, for desired heralding efficiencies close to unity the necessary increase in input power can be quite large. For a given desired heralding rate $J$, the required pump pulse energy for a given heralding efficiency $\eta$ is given by 
\begin{eqnarray}\label{E_pulse_required}
\mathcal{E}_\mathrm{pulse} = \frac{\hbar\omega_P\pi M}{2\Lambda\eta^2(1-\eta)}\sqrt{\frac{J}{\alpha f_P}}.
\end{eqnarray}
If a particular system is designed with $\Gamma=2M$ to maximize $J_\mathrm{heralds}$, which optimizes the \emph{power} efficiency of the device, it would suffer from a relatively poor heralding efficiency of $\eta=0.67$. If the same system were designed with $\Gamma\approx 100M$ to yield $\eta\approx 0.99$, using (\ref{E_pulse_required}) we find that the pulse energy to maintain the same heralding rate must increase by a factor of 15. The increased energy requirements, combined with the drastic reduction in the pulse duration needed to suppress spectral correlations, makes it a challenging task to design microresonator-based heralded single photon sources with high heralding efficiencies.

We have demonstrated that a tradeoff exists between the heralding rate and heralding efficiency for heralded single photon sources based on a \emph{single} microresonator source. Recent work has introduced the technique of spatial multiplexing \cite{Collins2013}, in which an array of $N\geq 2$ photon pair sources is used to generate independent single photons in multiple modes. These modes are fed into an optical switch, which routes the modes into a common output conditional on detection of the corresponding herald photon. It is important to note that such a technique, if designed using microresonator sources, does not evade the difficulties introduced by intraresonator losses. Denoting the individual heralding modes from each source by $i=1,\dots, N$ with corresponding rates of available photons $J_{\mathrm{HM}}^{(i)}$ and successful heralding rates $J_\mathrm{heralds}^{(i)}$, the overall heralding efficiency $\eta_\mathrm{tot}$ is given by
\begin{eqnarray}
\eta_\mathrm{tot}=\frac{\sum_{i=1}^{N}J_\mathrm{heralds}^{(i)}}{\sum_{i=1}^{N}J_\mathrm{HM}^{(i)}},
\end{eqnarray}
in which we have assumed perfect detection efficiency and neglected all losses except those in the microresonators. Assuming further that each source is characterized by the same quality factors, we find $\eta_\mathrm{tot}=\eta$, where $\eta$ is precisely the heralding efficiency of the individual sources. Nonetheless, multiplexing, while subject to the same requirements of over-coupling to attain high heralding efficiency, can be used to increase the overall heralding rate.

Our calculations demonstrate that loss has important consequences for microresonator-based heralded single photon sources. While arbitrary heralding efficiences and heralding rates -- up to the limits imposed by the minimum attainable pump pulse duration -- can be demanded from these sources, meeting such demands comes at the cost of increasing input power and decreasing pump pulse duration. As fabrication techniques improve and achievable intrinsic quality factors increase we expect the importance of these concerns to diminish; for the time being, however, the effects of loss must not be neglected.

\vspace{0.15in}
This work is supported by the Natural Sciences and Engineering Research Council of Canada and European COST Action MP 1403.

\bibliography{HeraldingDraft}
\end{document}